\documentstyle[aps,epsfig,multicol]{revtex}
\begin{document}

\title{A periodic elastic medium in which periodicity is relevant} 

\author{E.\ T.\ Sepp\"al\"a,$^*$  M.\ J.\ Alava,$^*$ and P.\ M.\ Duxbury$^\dag$}

\address{$^*$Helsinki University of Technology, Laboratory of Physics,\\ 
P.O.Box 1100, FIN-02015 HUT, Finland }
\address{ $^\dag$Dept. of Physics and Astronomy and Center for 
Fundamental Materials Research,\\
Michigan State Univ., E. Lansing, MI 48824-1116, U.S.A.}
\date{\today}

\maketitle

\begin{abstract}
We analyze, in both $(1+1)$- and $(2+1)$- dimensions,
a periodic elastic medium in which the periodicity is such
that at long distances the behavior is always 
in the random-substrate universality class.  
This contrasts with the models with an additive periodic
potential in which, according to the
field theoretic analysis of Bouchaud and Georges and more recently of 
Emig and Nattermann, the random manifold 
class dominates at long distances in $(1+1)$- and $(2+1)$-dimensions. 
 The models we use are random-bond Ising
interfaces in hypercubic lattices.  The exchange constants are random
in a slab of size $L^{d-1} \times \lambda$ and these coupling
constants are periodically repeated along either $\{10\}$ or $\{11\}$
(in $(1+1)$-dimensions) and $\{100\}$ or $\{111\}$ (in
$(2+1)$-dimensions).  Exact ground-state calculations 
confirm scaling arguments which predict that the surface roughness $w$
behaves as: $w \sim L^{2/3}, \ L \ll L_c$ and $w \sim L^{1/2},\ L \gg
L_c$, with $L_c \sim \lambda^{3/2}$ in $(1+1)$-dimensions and; $w \sim
L^{0.42}, \ L \ll L_c$ and $w \sim \ln(L), L \gg L_c$, with $L_c \sim
\lambda^{2.38}$ in $(2+1)$-dimensions.
\end{abstract}

\noindent {\it PACS \# \ \ 05.70.Np, 75.10.Nr, 02.60.Pn, 68.35.Ct}

\begin{multicols}{2}[]
\narrowtext

\section{Introduction}

Periodic elastic media arise in a surprising array of problems,
including spin or charge density waves; flux line lattices; and random
magnets.  A model frequently used\cite{Bouchaud92,Emig98,Emig99} to
describe a manifold, defined by the single valued height variable
$h(\vec{r})$ in a periodic elastic medium is
\begin{equation}
{\mathcal H}_{pem} = \int d\vec{r} \left\{ {\gamma\over 2} 
[\nabla h(\vec{r})]^2 + \eta[h(\vec{r})] + V_p[h(\vec{r})] \right\},
\label{eq_pem}
\end{equation}
where $V_p$ is a periodic potential in the height direction and the random
potential $\eta$ is not periodic.  This is directly analogous to the
model used to study lattice effects in thermal roughening and in field
theoretic studies of commensurate phases in Ising magnets with
competing interactions.  In the model (\ref{eq_pem}), the periodic
potential is non-random and tends to pin the interface while the
quenched random pinning $\eta[h(\vec{r})]$ prefers to make the
interface wander.  The surface tension term ${\gamma\over 2} [\nabla
h(\vec{r})]^2$ seeks a flat interface and also competes with the
quenched random pinning.  Field theoretic
calculations\cite{Bouchaud92,Emig98,Emig99} suggest that at long
distances, for (1+1)- and (2+1)- dimensional interfaces, the periodic
pinning potential is irrelevant, and hence that the interface scaling
behavior is in the random-bond-Ising universality class where width
$w^2 = \langle h^2 \rangle - \langle h \rangle^2 \sim L^{2 \zeta}$
with the roughness exponent $\zeta = 2/3$ in $(1+1)$, and $\zeta \approx
0.21(4-d)$ in $(d+1),\ d\ge 2$
\cite{Huse85a,Huse85b,Fisher86,Middleton95,Alava96}.  
Note that lattice calculations are strongly affected by a 
lattice pinning potential and have a flat phase even for
large lattice sizes \cite{Alava96}.
 
Another problem which has been heavily studied is the random substrate
problem\cite{Cardy82,Toner90,Shapir92}.  This was introduced to model
the effect of a random substrate on layers of absorbed atoms, and also
serves as a model for the effect of a p-fold random field on the
XY-model\cite{Cardy82}.  There is now a consensus that there is a
disorder dominated glassy phase in this model (in 2 substrate
dimensions) at low temperatures that is reflected in long distance
correlations which behave as $C(r) \sim \ln^2|r|$ (in contrast to
thermally rough correlations in dimension $(2+1)$, which grow as as
$C(r) \sim \ln|r|$).  There has been some uncertainty about whether
the leading order correlations found by Cardy and Ostlund
(CO)\cite{Cardy82} are correct, with functional renormalization group
calculations agreeing with CO \cite{Toner90,Shapir92}, and variational
calculations disagreeing.  The substrate roughness is randomly drawn
from the interval $(0,1)$ (in lattice units).  This corresponds to a
different sort of periodic elastic medium than that described in
(\ref{eq_pem}) above.  Here, the random substrate leads to a
periodically repeated disorder seen by an interface lying above the
random substrate.  This arises due to the fact that the first, third,
fifth, etc. atoms deposited at the same position on the random
substrate see exactly the same disorder when they land.  This
corresponds to a random-bond Ising magnet in which the disorder is
repeated with period $\lambda=2$ along the growth direction.  In
general, the disorder may range over a scale $(0,\lambda-1)$, and this
leads to a periodic variation in the disorder on length scale
$\lambda$.  The continuum model for this system is simply,
\begin{equation}
{\mathcal H}_{p} = \int d \vec{r} \left\{ {\gamma\over 2} 
[\nabla h(\vec{r})]^2 + \eta[h(\vec{r})] \right\},
\label{eq_p}
\end{equation}
but where $\eta$ is {\it periodic in $h(\vec{r})$}, so that we require
$\eta[h(\vec{r})+\lambda] = \eta[h(\vec{r})]$.  There has been
considerable study of the random substrate ($\lambda=2$) problem, with
the early controversy now being resolved in favor of a ``super-rough''
``Bragg-glass'' phase in $(2+1)$-dimensions in which $w\sim \ln(L)$.
Exact ground-state calculations have been very useful in resolving
this controversy \cite{Rieger95,Zeng96,Rieger97,Zeng98}.  It is quite
easy to see (see Section III) that in $(1+1)$-dimensions, the random
substrate problem behaves as a random walk (RW), so that $w\sim L^{1/2}$.
Note however it has been recently argued that although typical
dislocations do not destroy the ``Bragg-glass'' ground state, optimal
dislocations have negative energy, and hence are expected to destroy
the Bragg glass in $(2+1)$-dimensions\cite{Zeng99,Pfeiffer00}.

In this paper we study the Hamiltonian (\ref{eq_p}) as a function of
the periodicity $\lambda$ of the disorder.  We show that at long
length scales in $(1+1)$- and $(2+1)$-dimensions, the periodicity is
relevant and that the random substrate universality class holds.  The
paper is arranged as follows: Section \ref{model} sets up the model and
describes the way in which we calculate the exact positions of
interfaces in random Ising magnets.  The scaling theory describing the
behavior of these interfaces is developed and tested in Section \ref{scaling}.
We give a brief conclusion in Section \ref{concl}.

\section{Discrete model and exact algorithm}
\label{model}

The model which we use to analyze the effect of 
periodic disorder on interface properties is a spin-half Ising 
system with random bonds (RB) on square and cubic lattices. 
 The Hamiltonian is given by, 
\begin{equation}
{\mathcal H}_{RB} = -\sum_{\langle ij \rangle} J_{ij} S_i S_j,
\label{HRB}
\end{equation} 
where $J_{ij}>0$ are coupling constants and the spin variables $S_i$
take the values $\pm 1$.  The spins on two opposite boundaries of the
lattices, $z=1$ and $z=L$, are fixed and have opposite signs so that
an interface must exist in the lattice.  Our calculations are at zero
temperature and we find the ground-state interface properties for
interfaces whose average normals lie in the $\{10\}$ or $\{11\}$
directions of square lattices and in the $\{100\}$ or $\{111\}$
directions of cubic lattices.  The coupling constants are random in a
slab of size $L^{d-1} \times \lambda$ and then periodically repeated
$L/\lambda$ times along a chosen direction. The distributions
used for the $J_{ij}$'s vary here from case to case but
are always chosen so that the interfaces are
rough even for small lattices sizes, and even in the
$\{100\}$ orientation cubic systems. In Fig.\ \ref{fig1}, we illustrate the way in which the
periodic disorder is implemented for the $\{10\}$ and $\{11\}$
directions of a square lattice.  As is now well known
\cite{Middleton95,Alava96,Alava00}, the ground state interface of the
system (\ref{HRB}) can be found {\it exactly} using the maximum flow
algorithm.  We have a custom implementation of the push-relabel
algorithm for this problem and using it we are able to find the exact
ground state interface in Ising systems of size one millions sites in
about one minute of CPU time on a high end workstation.

\section{Scaling theory and numerical results}
\label{scaling}

\begin{figure}[f]
\centerline{\epsfig{file=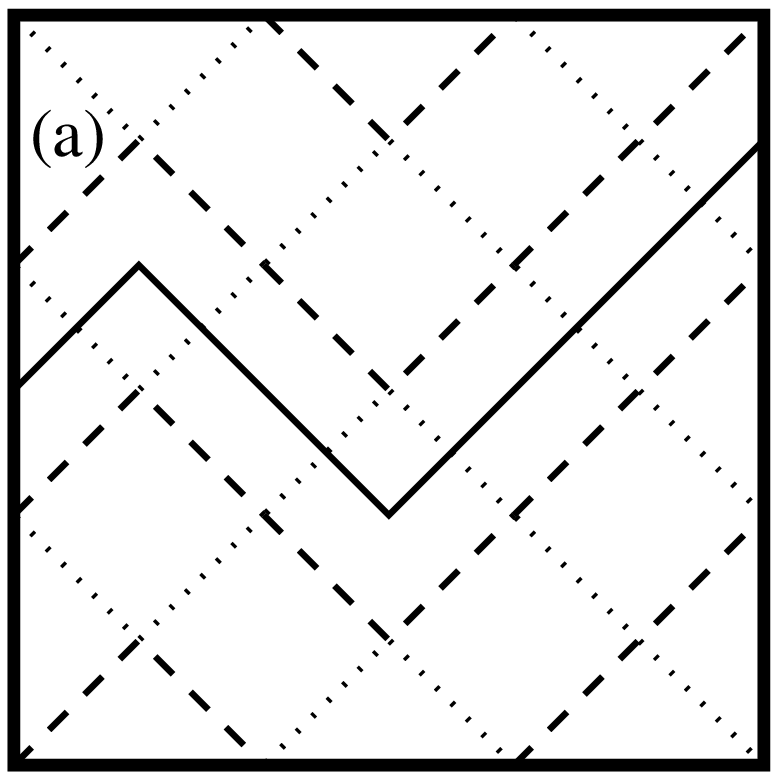,width=6cm}}
\centerline{\epsfig{file=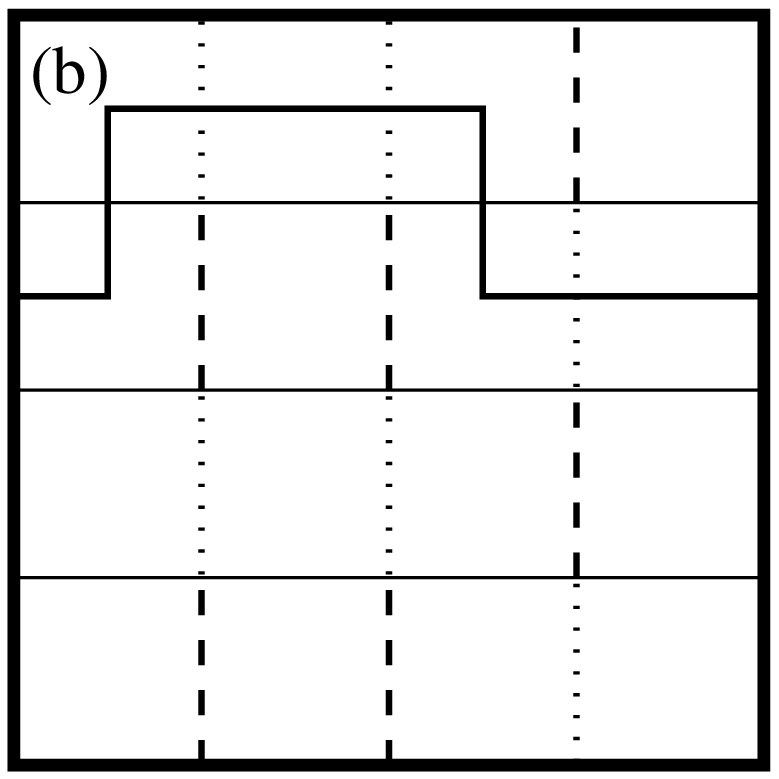,width=6cm}}
\caption{An example of interface in a random substrate 
problem, with period $\lambda=2$: (a) in $\{11\}$ orientation; (b) the
$\{10\}$ orientation.  The dotted lines, $\cdots$, describes the lower
energy bond of the two bonds (in the system of period 2), while the
dashed line, --~--, describes the higher energy bond. A minimum energy
path through each system is indicated with a thick solid line. }
\label{fig1}
\end{figure}

Consider the ground-state interface of a square lattice in which the
bond disorder has period two in the $\{11\}$ orientation
(e.g., Fig. \ref{fig1}(a)).  It is obvious that the interface is highly
degenerate, as the ground state interface may start in any of $L/2$
equivalent positions. Consider now starting to create a ground state
interface from the left side of Fig.\ \ref{fig1}(a).  To minimize the
interface energy one chooses the weakest bond.  Having chosen this
weakest bond, the interface crosses this weakest bond and chooses the
weakest bond in the next column.  This process of choosing the weakest
bond continues across the sample and, for period two, the random walk
so generated gives the {\it exact} ground state.  The reason this
ground state is exact is that at each step, all of the possible random
bonds in each column are tested (there are only two!).  Thus in this
limit, $w\sim L^{1/2}$ as for a random walk.  In contrast, if the
period diverges, the model returns to the random bond Ising
universality class (or equivalently the directed polymer (DP) in a random
medium) for which $w\sim L^{2/3}$.  For finite $\lambda$, we expect
that the interface will seek to optimize its global wandering until
the roughness reaches the wavelength of the
periodicity~\cite{comm}. After that it has exhausted all possibilities
and then returns to a random walk behavior.  We thus have,
\begin{equation}
w(L,\lambda) \sim \left\{ \begin{array} {lll}
L^{2/3},&\mbox{\hspace{5mm}}&w \ll \lambda, \\
L^{1/2},& &w \gg \lambda. 
\end{array} \right.
\label{DPRW} 
\end{equation}
A natural scaling form based on these limiting behaviors is,
\begin{equation}
w(L,\lambda) \sim L^{2/3} f\left(\frac{L}{\lambda^{3/2}}\right).
\label{scalingDP} 
\end{equation}
where the scaling function $f(z)$ for the roughness has the asymptotic
behavior,
\begin{equation}
f(z) \sim \left\{ \begin{array} {lll}
\mbox{const} &\mbox{\hspace{5mm}}& z \ll 1, \\
z^{-{1/6}} & & z \gg 1.
\end{array}\right. 
\label{2dscalingf} 
\end{equation}

\begin{figure}[f]
\centerline{\epsfig{file=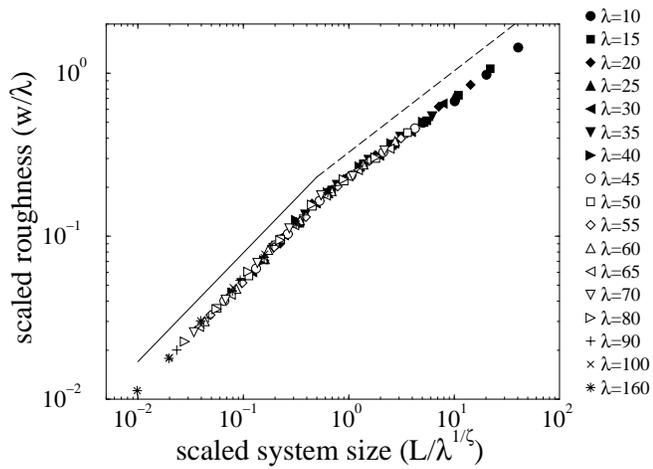,width=65mm,angle=-90}}
\caption{(a) The roughness ($w$) of manifolds divided by the 
wavelength of the periodicity ($\lambda$) vs.\ normalized system size
($L/\lambda^{1/\zeta}$), where $\zeta = \zeta_{DP} = 2/3$, for
$\{10\}$ oriented $(1+1)$-dimensional systems. The random
bonds are from a uniform distribution with strength $\Delta
J_{ij,\perp}/J_0 =1$ in the perpendicular ($z$) direction, and $\Delta
J_{ij,\|}/J_{0,\|} =0.1$ in the parallel ($x$) direction in all layers in
order to break the degeneracy.  $J_{0,\|}/J_0 =0.2$. The number of
realizations $N=200$ for each wavelength $\lambda \in [10,..,160]$ and
system size $L^2 \in [20^2-1280^2]$.  The solid line, ---, has a slope
$\zeta = \zeta_{DP} = 2/3$ and the dashed line, --~--, has a slope
$\zeta = \zeta_{RW} = 1/2$.}
\label{fig2}
\end{figure}

\begin{figure}[f]
\centerline{\epsfig{file=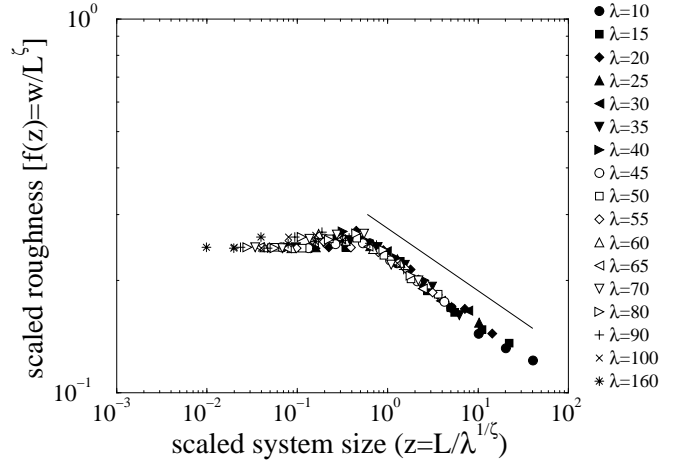,width=65mm,angle=-90}}
\caption{The scaling function $f(z) = w/L^{\zeta}$ of the 
roughness $w(L,\lambda)$ vs.\ scaling parameter
$z=L/\lambda^{1/\zeta}$, where $\zeta = \zeta_{DP} = 2/3$ for the same
data as in Fig.~\ref{fig2}.  The solid line has a slope of
$\zeta_{RW} - \zeta_{DP} = -1/6$.}
\label{fig3}
\end{figure}

Tests of the asymptotic behaviors (\ref{DPRW}) and the scaling
function (\ref{scalingDP}) and the results are presented in Figs.\
\ref{fig2} and \ref{fig3} for the $\{10\}$ orientation.  It is seen
that the predictions of the scaling theory are nicely confirmed.
Similar results were found for the $\{11\}$ orientation, too.

\begin{figure}[f]
\centerline{\epsfig{file=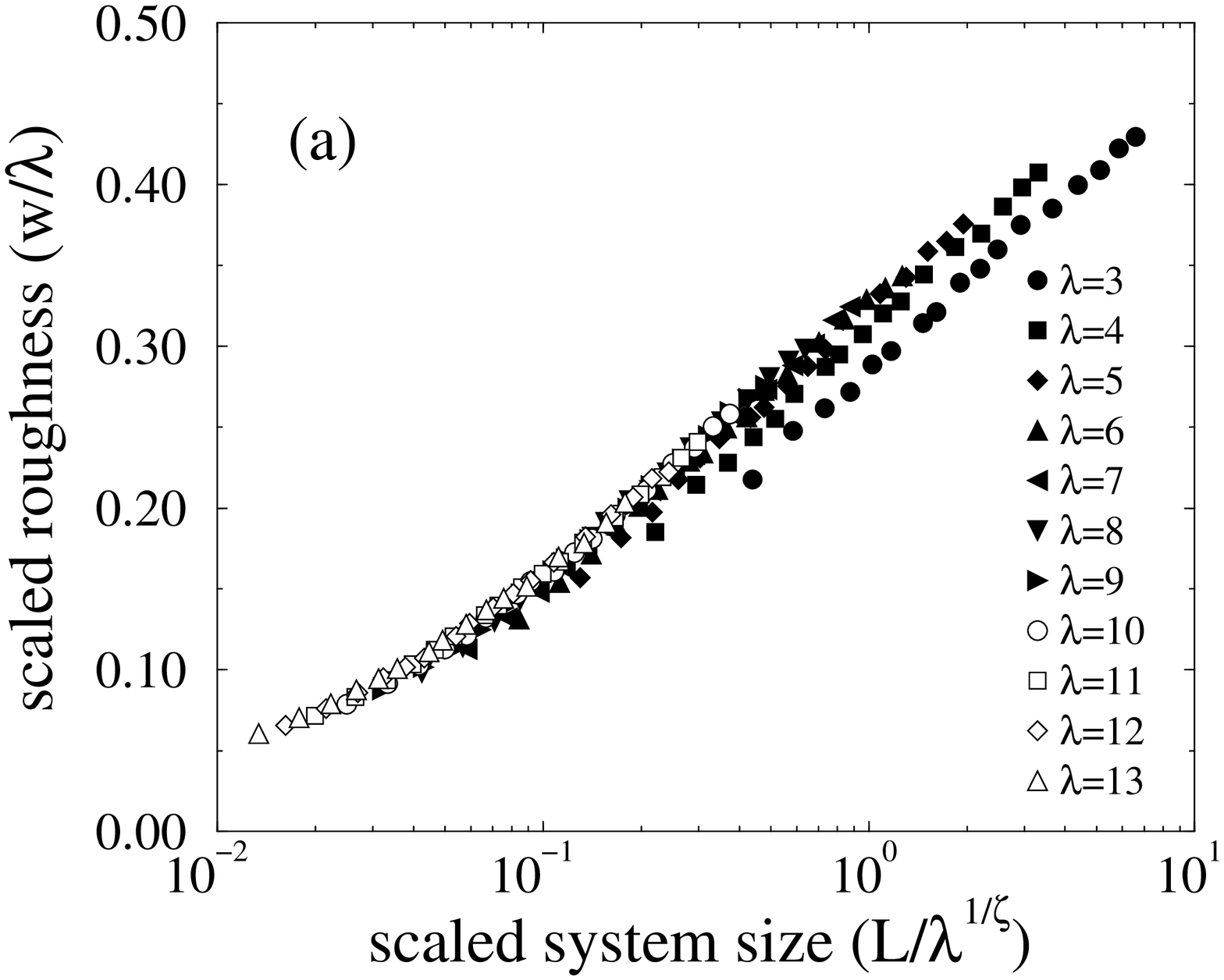,width=65mm,angle=-90}}
\centerline{\epsfig{file=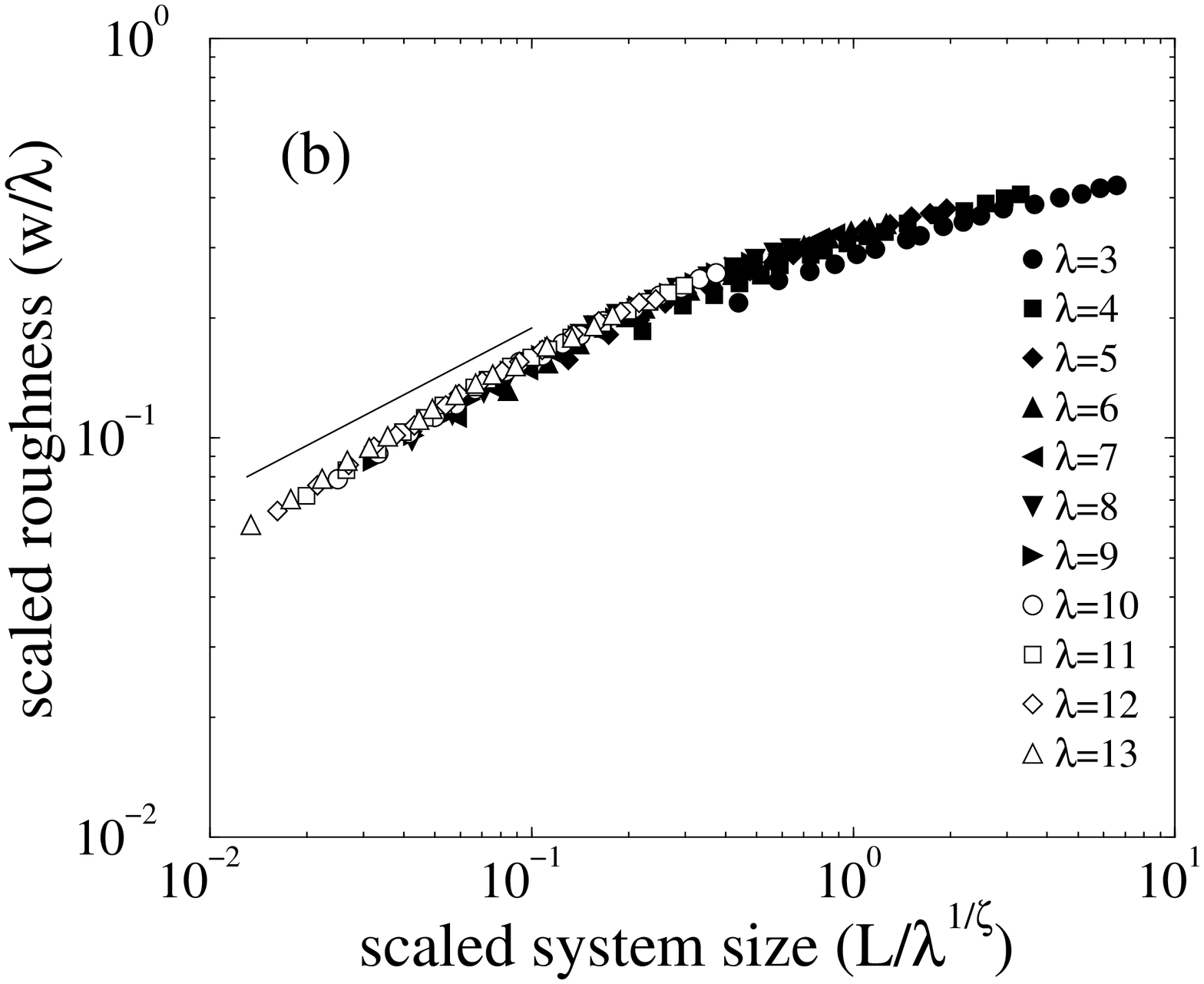,width=65mm,angle=-90}}
\centerline{\epsfig{file=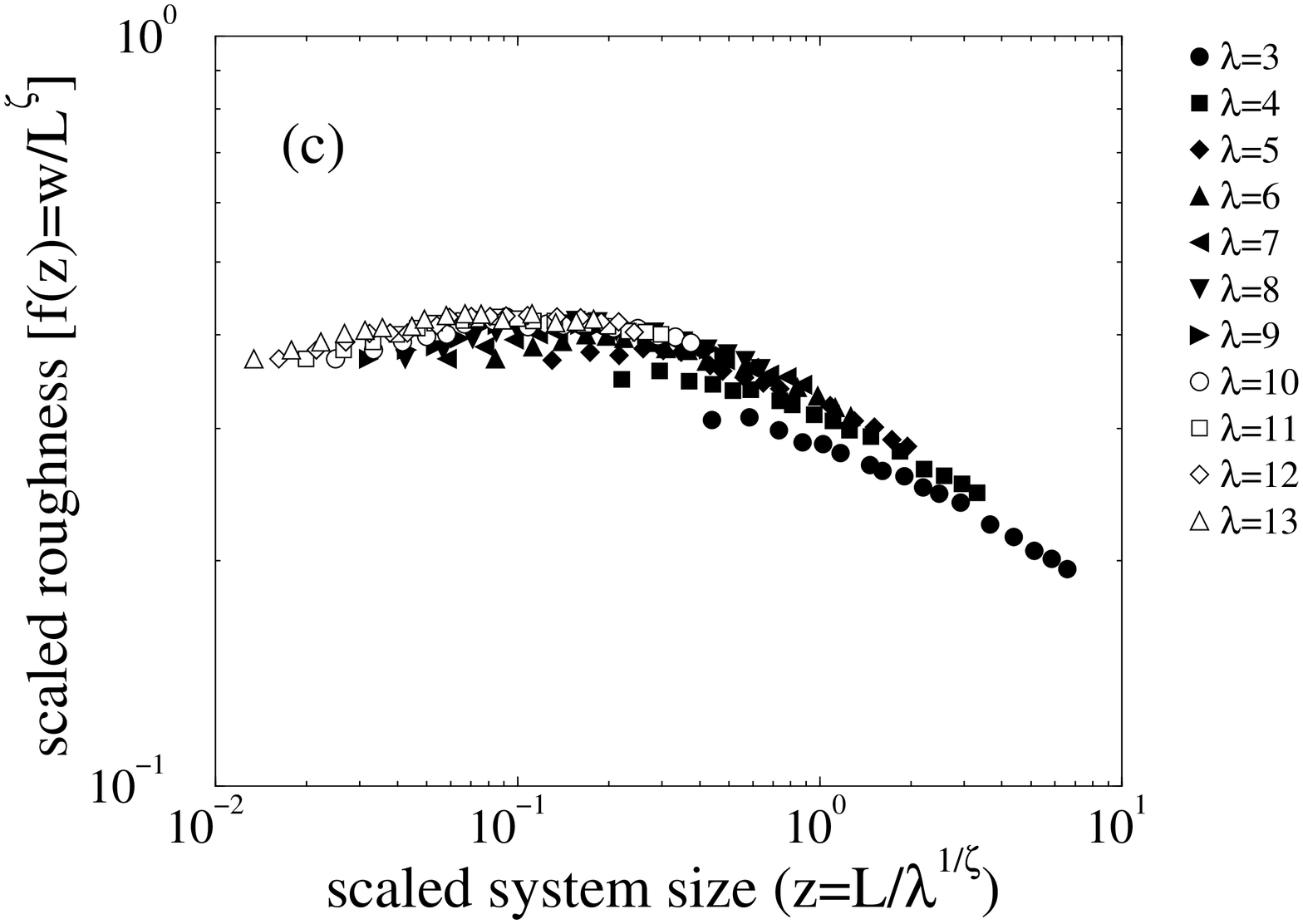,width=65mm,angle=-90}}
\caption{(a) and (b) The data-collapse, $w/\lambda$ vs. $L/\lambda^{1/\zeta}$,
$\zeta = \zeta_{RB} =$ $0.42$, for the roughness of
$(2+1)$-dimensional $\{ 100 \}$ oriented systems.  The random bonds
are from uniform distribution with $\Delta J_{ij,\perp}/J_0 =1$ in the
perpendicular ($z$) direction and constant $J_{ij,\|}/J_0 =0.2$ in the
parallel ($x,y$) direction.  The number of realizations $N=200$ for
each wavelength $\lambda \in [3,..,13]$ and system size $L^3 \in
[6^3-90^3]$.  (c) The scaling function $f(z) = w/L^{\zeta}$ of the
roughness $w(L,\lambda)$ vs. scaling parameter
$z=L/\lambda^{1/\zeta}$.  Finite-size effects with logarithmic
corrections are visible as a curvature for small $L$.}
\label{fig4}
\end{figure}

We turn now to the behavior of random surfaces in $(2+1)$ dimensions.
There, renormalization group (RG) techniques have been applied to the
random-phase sine-Gordon model~\cite{Cardy82,Toner90,Shapir92}, to
random bond interfaces and to fairly general models of periodic elastic
media.  Numerically,  exact maximum-flow-minimum-cut and
minimum-cost-matching algorithms~\cite{Zeng96} and Monte Carlo methods
\cite{Cule95}) have been used.  In the random substrate problem, there
is a low temperature ``super-rough'' phase where $w^2 \sim
\ln^2(L)$, while in the random manifold problem, the
surface roughness is found to behave as $w \sim L^{\zeta_{RB}}$,
where $\zeta_{RB} = 0.42\pm 0.01$. 
The qualitative reasoning expressed in the first paragraph
of this section also applies to higher dimensions,
so that we expect the behavior of ${\mathcal H}_p$ to be in the
random substrate universality classes at long length
scales $w>\lambda$, while the random manifold universality
class is dominant at short length scales $w < \lambda$.
The limiting behaviors in dimension (2+1) are then,
\begin{equation}
w(L,\lambda) \sim \left\{ \begin{array} {lll}
L^{\zeta_{RB}},&\mbox{\hspace{5mm}}&w \ll \lambda , \\
\ln L,& &w \gg \lambda . 
\end{array} \right.
\label{RBlog} 
\end{equation}
We thus expect,
\begin{equation}
w(L,\lambda) \sim L^{\zeta_{RB}}f\left(\frac{L}{\lambda^{1/\zeta_{RB}}}
\right),
\label{scalingRB} 
\end{equation}
and that the scaling function in $(2+1)$-dimensions is
\begin{equation}
f(z) \sim \left\{ \begin{array} {lll}
\mbox{const} &\mbox{\hspace{5mm}}& z \ll 1, \\
\ln z/z^{\zeta_{RB}} & & z \gg 1,
\end{array}\right. 
\label{3dscalingf} 
\end{equation}
with the scaling parameter $z=L/\lambda^{1/\zeta_{RB}}$.  The
asymptotic behaviors of Eq.~(\ref{RBlog}) are illustrated in Figs.\
\ref{fig4}(a) and (b) for interfaces in the $\{100\}$ orientation.
The logarithmic asymptotic behavior is clearly confirmed in Fig.\
\ref{fig4}(a), but the random manifold behavior is still strongly
effected by finite size effects.  This is understandable as large
system sizes are necessary to see the asymptotic random manifold
behavior, even in the $\lambda \to \infty$ limit
\cite{Alava96,Middleton95}.  Though finite size effects are clearly
evident in the scaling plot of Fig.\ \ref{fig4}(c), the data collapse
at large $\lambda$ is quite satisfying.  It is clear that the random
substrate (Bragg glass) universality class\cite{Rieger95,Zeng96} is
dominant at large enough length scales.  We have tested the behavior
in the $\{111\}$ orientations and find that $\{111\}$ interfaces
behave in a similar manner.

\section{Conclusions}
\label{concl}

We have studied the scaling behavior of an elastic manifold in the
presence of a periodically repeated ``strong'' bond disorder.  
We find that in $(1+1)$- and in
$(2+1)$- dimensions, and at long distances, the periodicity is
relevant so these interfaces are in the random substrate universality
class.  This is to be contrasted with an interface in a system with a
periodic potential and with random disorder.  In the latter problem
the periodic potential is claimed to be irrelevant on long length
scales in $(1+1)$- and $(2+1)$-dimensions for any
disorder\cite{Bouchaud92,Emig98,Emig99}, though at weak disorder
 numerical work on
$\{100\}$ orientation cubic lattices indicate a strong tendency to
order due to lattice effects\cite{Alava96,Raisanen98}.

ETS and MJA thank the Academy of Finland for financial support.  PMD
thanks the DOE under contract DOE-FG02-090-ER45418 for support.


\newpage

\end{multicols}
\widetext


\begin{references}
\bibitem{comm} In fact, the cross-over roughness value, when the
interface starts to feel the periodicity is $w_*/\lambda \simeq 0.25
\pm 0.02$ in $(1+1)$ dimensions.
\end{references}
\end{document}